# Effect of evolutionary physical constants on type-1a supernova luminosity

Rajendra P. Gupta*

*Department of Physics, University of Ottawa, Ottawa, Canada K1N 6N5*

**ABSTRACT**

Type 1a supernovae, SNeIa, are used as standard candles in cosmology for determining the distances of the galaxies harboring them. We show that the luminosity of an SNIa depends on its distance from us when physical constants (the speed of light $c$, the gravitational constant $G$, and the Planck constant $h$) are permitted to evolve. It is because the Chandrasekhar mass of the white dwarf that explodes to create SNIa depends on the values of the constants at the epoch the SNIa is formed. We show that the SNeIa luminosities were up to about four times higher in the past than they are now. Thus, the luminosity distance estimation of the earliest SNeIa could be off by up to a factor of two. Cosmological parameters, determined with this correction applied to the redshift vs. distance modulus database (Pantheon SNeIa), are not very different from those from the standard ΛCDM model without this correction, except for the dark-energy density and the curvature energy density; the latter increases at the cost of the former. Variations of the constants are given by $\dot{G}/G = 3.90(\pm 0.04) \times 10^{-10}\ yr^{-1}$ and $\dot{c}/c = \dot{h}/h = 1.30(\pm 0.01) \times 10^{-10}\ yr^{-1}$ at present. These variations are valid only when $G$, $c$, and $h$ are permitted to vary concurrently rather than individually.

**Key words:**

## 1  INTRODUCTION

Determination of distances of cosmological objects is a challenging task. Type 1a supernovae, SNeIa, have proven to be the most reliable standard candle for measuring distances of such objects. Their brightness is comparable to that of the host galaxy for several weeks, and thus they are visible from vast distances. The redshift vs. luminosity observations of SNeIa led to the discovery of the accelerated expansion of the Universe. When a white-dwarf star in a close binary accretes enough mass from its companion to reach Chandrasekhar mass, $M_{ch}$, it undergoes cataclysmic explosion due to runaway thermonuclear fusion. Since $M_{ch}$ is involved in all such explosions, the peak brightness and time it takes to reach the peak brightness are very well correlated. They, in turn, are used to standardize the luminosity of the event, which has been dubbed type 1a supernova in contrast to other supernovae events, which cannot be standardized.

The SNIa standard candle is based on $M_{ch}$ to be the same at all cosmological times. Since $M_{ch} \propto (hc/G)^{3/2}$, the evolution of any of the constants will affect $M_{ch}$, and hence the SNeIa luminosity will also become evolutionary; here $h$ is Planck constant, $c$ is the speed of light, and $G$ is the gravitational constant. Thus, one may, in principle, be able to constrain the variation of $hc/G$ from the analysis of the redshift vs. luminosity observations data, such as the 1048 SNeIa Pantheon data (Scolnic et al. 2018). However, such studies are strongly model-dependent; if a model is developed based on fitting the data assuming $hc/G$ to be constant, then the same model *and* parameters cannot be used to constrain $hc/G$. Another critical point to note is that it may not be prudent to assume only $G$ to be varying since this would imply that we know for sure that $h$ and $c$ are not cosmologically evolutionary.

There has been a great interest in determining the evolution of physical constants, especially Newton's gravitational constant $G$, since Dirac (1937) predicted its variation based on his large number hypothesis. Teller (1948) was the first to suggest a constraint on the variation of $G$ from the stellar scaling laws applied to the evolution of Solar luminosity and the environment required for the existence of life on Earth in the past. Since then, many methods have been developed to determine the variation of $G$, which have all resulted in the constraints on $\dot{G}/G$, well below that predicted by Dirac. These include methods based on solar evolution (Teller 1948, Chin & Stothers 1976, Sahini & Shtanov 2014), lunar occultation and eclipses (Morrison 1973), paleontological evidence (Sisterna & Vucetich), white dwarf cooling and pulsation (Benvenuto et al. 1999, Garcia-Berro et al. 2011, Corsico et al. 2013), star cluster evolution (Degl'Innocenti 1995), neutron star masses and ages (Thorsett 1996), cosmic microwave background temperature anisotropies (Bai et al. 2015, Ooba et al. 2017), big-bang nucleosynthesis abundances (Copi et al. 2004, Alvey et al. 2020), asteroseismology (Bellinger & Christensen-Dalsgaard 2019), lunar laser ranging (Williams et al. 2004, Hofmann and Müller 2018), the evolution of planetary orbits (Pitzeva & Pitzev 2013, Fienga et al. 2014, Genova et al. 2018), binary pulsars (Damour et al. 1988, Kaspi et al. 1994, Zhu et al. 2019), and supernovae type-1a luminosity evolution (Gaztañaga et al. 2001, Wright & Li 2018).

While Einstein developed his ground-breaking theory of special relativity based on the constancy of the speed of light, he did consider its possible variation (Einstein 1907). This was followed by the varying speed of light theories by Dicke (1957), Petit (1988), and Moffatt (1993a, b). More recently, Albrecht & Magueijo (1999) and Barrow (1999) developed such a theory in which Lorentz invariance is broken as there is a preferred frame in which scalar field is minimally coupled to gravity. Other proposals include locally invariant theories (Avelino & Martins 1999; Avelino, Martins & Rocha 2000) and vector field theories that cause spontaneous violation of Lorentz invariance (Moffat 2016).

It is a standard practice in physics to vary one parameter while keeping others constant in laboratory experiments since one can control the parameters. Applying this approach to astrophysical problems is not possible as we cannot control any parameter. We cannot treat the variation of one parameter - a physical constant in our context - while ignoring the potential variation of others in analyzing any problem. It is also a standard practice to apply local energy conservation laws to the problems involving evolutions at a cosmological time scale when it is well known that energy is not conserved in general relativity and cosmology (Harrison 1981, Peebles 1993, Baryshev 2008, Carroll 2010, Weiss & Baez 2017, Filippini 2020, Velten and Caramês 2021). Any conclusion reached due to these standard practices should thus be treated with caution and not considered proof against Dirac's prediction. Our approach has been to deviate from these traditional practices to analyze various problems that are fundamentally cosmological (Gupta 2020, 2021a-c, 2022).

We have developed a relativistically covariant cosmological model (Gupta 2020), which permits the variation with time of $G$, the speed of light $c$, the Planck constant $h$, and the Boltzmann constant $k_B$. With this variable physical constants (VPC) model, we have shown that the most reliable and stringent constraints on $\dot{G}/G$ determined from orbital timing of bodies in the solar system (Williams et al. 2004, Hofmann and Müller 2018) and binary pulsars (Damour et al. 1988, Nordtvedt 1990, Kaspi et al. 1994, Zhu et al. 2019) are not on $\dot{G}/G$ but on $(\dot{G}/G - 3\,\dot{c}/c)$ (Gupta 2021a). Thus, if one ignores the variation of $c$, the constrain determined is on the variation of $G$. Using the same model, we were able to (a) resolve the primordial lithium problem from the Big-Bang nucleosynthesis (Gupta 2021b); (b) establish that $\dot{c}/c$ cannot be constrained from gravitational lensing (Gupta 2021c); and (c) provide a reasonable solution to the faint young Sun problem (Gupta 2022).

Here we analyze the basis of SNeIa luminosity evolution in the papers of Gaztañaga et al. (2001) and Wright and Li (2018) and show that the constraints they determined on $\dot{G}/G$ are valid only when the potential variations of other constants, such as the speed of light $c$ and the Planck's constant $h$ are ignored. The approach is to first determine the evolution of SNeIa luminosity in the VPC scenario (Section 2). Then we see how it affects luminosity distance determination and thus the distance modulus (Section 3). In Section 4, we use the findings of Section 3 to fit the Pantheon SNeIa data to determine key cosmological parameters and compare the quality of fit and data with that of the standard ΛCDM model. We discuss our findings in Section 5 and present the conclusions in Section 6.

## 2 SUPERNOVAE TYPE-1A LUMINOSITY EVOLUTION

The peak SNIa luminosity, $L_{SN}$, is proportional to the mass of the nickel synthesized in the white dwarf explosion resulting in the SNIa, which is proportional to Chandrasekhar mass $M_{Ch}$ of the exploding mass (Arnett 1982, Gaztañaga et al. 2001, Wright & Li 2018). The explosion energy is partially used up to counter the gravitational binding energy, $E_{gr}$, of the star and the balance is converted into the kinetic energy, $E_{ke}$. A fraction of this kinetic energy is radiated out and observed as SNIa luminosity. Therefore,

$$\eta M_{Ch} c^2 \equiv E_{kin} + E_{gr}. \tag{1}$$

Here, $\eta$ represents the efficiency of mass to energy conversion. Now, the Chandrasekhar mass and the radius $r_{wd}$ of a white dwarf are given by (Maoz 2016),

$$M_{ch} = 0.21 \left(\frac{Z}{A}\right)^2 \left(\frac{hc}{Gm_p^2}\right)^{3/2} m_p, \text{ and} \quad (2)$$

$$r_{wd} \approx \frac{h^2}{10 m_e m_p^{5/3} G} \left(\frac{Z}{A}\right)^{5/3} M_{ch}^{-1/3}. \quad (3)$$

Here $Z$ is the atomic number, $A$ is the atomic mass number, $m_p$ is the proton mass, and $m_e$ is the electron mass. The gravitational binding energy is (e.g., Maoz 2016)

$$E_{gr} \approx -\frac{GM_{ch}^2}{r_{wd}}. \quad (4)$$

Let us follow closely the VPC model (Gupta 2020), which defines the variation of $c, G,$ and $h$ as a function of the scale factor $a$ as

$$c(a) = c_0 f(a); G(a) = G_0 f(a)^3; h(a) = h_0 f(a), \quad (5)$$

with $f(a) = \exp(a^\alpha - 1)$. Here subscript 0 indicates the quantity as measured by the observer at present $t = t_0$, i.e., at $a = 1$, and $\alpha$ is constant determined analytically (Gupta 2018) and confirmed by the best data fit. We may now write the following scaling relations for the white dwarf of Chandrasekhar mass $M_{Ch}$:

$$M_{Ch} c^2 \sim \left(\frac{f \times f}{f^3}\right)^{3/2} f^2 = f^{1/2}, \quad (6)$$

$$r_{wd} \sim \frac{f^2}{f^3} \times \left(f^{-3/2}\right)^{-1/3} = f^{-1} \times f^{1/2} = f^{-1/2}, \text{ and} \quad (7)$$

$$E_{gr} \sim \frac{GM_{Ch}^2}{r_{wd}} \sim \left(\frac{f^3 \times f^{-3}}{f^{-1/2}}\right) = f^{1/2}. \quad (8)$$

Since $M_{Ch}c^2$ and $E_{gr}$ both are scaling as $f^{1/2}$ in equation (1), we may write the scaling of the energy, $E_{SN}$, contributing to the SNIa luminosity as

$$E_{SN} \propto E_{kin} = (\eta M_{Ch} c^2 - E_{gr}) \sim f^{1/2}. \quad (9)$$

We also need to consider how the energy of atomic energy levels involved in the emission of radiation, which affects the luminosity, scales when the coupling constants vary. The energy levels are proportional to the Rydberg unit of energy: $R_y = hcR_\infty$ where $R_\infty = m_e e^4 / 8\epsilon_0^2 h^3 c$ is the Rydberg constant. Here $e$ is the electron charge and $\epsilon_0$ is the permittivity of space given by $c = 1/\sqrt{\mu_0 \epsilon_0}$ with $\mu_0 = 4\pi \times 10^{-7}$ henry per meter the permeability of free space. All masses, electric charges, and permeability are constant in the VPC model. This means,

$$R_\infty \sim c^4/(h^3 c) \sim f^0, \text{ and} \quad (10)$$

$$R_y \sim f^2 \times f^0 \sim f^2. \quad (11)$$

Thus, the number of photons, $N$, released in the SNIa explosion

$$N \propto E_{SN}/R_Y \sim f^{1/2}/f^2 \sim f^{-3/2} \Rightarrow N = N_0 f^{-3/2}. \quad (12)$$

Consider now the evolution of the photon energy itself. It is given by $ch/\lambda$ where $\lambda$ is the photon wavelength. The lower energy of a photon at the time of emission by a factor $f^2$ due to the lower values of $c$ and $h$ is offset by the increase in the photon energy at the time of its detection due to the higher value of $c$ and $h$ by a factor $f^{-2}$ for a given photon wavelength $\lambda$. However, the photon wavelength does expand due to the expansion of the Universe as the scale factor $a$, and this must be taken into account in calculating the detected photon energy and luminosity.

## 3 LUMINOSITY DISTANCE AND REDSHIFT

Following Gupta (2020), we may write the photons emitted by a source at the time $t_e$ are spread over a sphere of radius $S_k(d_P)$ and area $A_P(t_0)$ by the time photons reach the observer at the time $t_0$. Here $d_P$ is the proper distance of the source from the observer, $S_k(d_P) = R \sin(d_P/R)$ for $k = +1$ (closed Universe); $S_k(d_P) = d_P$ for $k = 0$ (flat Universe), $S_k(d_P) = R \sinh(d_P/R)$ for $k = -1$ (open Universe), where $R$ is the parameter related to the curvature of the Universe. We may write the area of the sphere

$$A_P(t_0) = 4\pi S_k(d_P)^2. \quad (13)$$

The photon energy flux is defined as luminosity $L$ divided by the area in a stationary universe. When the Universe is expanding, then the flux is reduced by a factor $1 + z \; (\equiv a^{-1}$, with $z$ being the redshift) due to energy reduction of the photons from the change in their wavelength:

$$\lambda_0 a = \lambda_e \Rightarrow \lambda_0 = (1+z)\lambda_e.$$

The photon energy is thus altered by a factor of $1/(1+z)$ due to the expanding Universe.

We also need to determine how the increase in the time interval of the emitted photons affects the flux. The proper distance between two emitted photons separated by a time interval $\delta t_e$ is $c_e \delta t_e$ whereas the proper distance between the same two photons when detected by observation is $(c_e \delta t_e)(1+z)$, and the time interval between the same two photons is $\delta t_0 = (c_e \delta t_e)(1+z)/c_0 = (1+z)f\delta t_e$. Thus, the time duration between the photons has altered by a factor of $\delta t_e/\delta t_0 = 1/[f(1+z)]$, which is time dilation, that we have to consider in estimating the flux, and therefore in calculating the source distance.

The above two effects alter the photon energy flux cumulatively by a factor $1/[f(1+z)^2]$. We need to consider now the effect of the change in the number of photons emitted as per equation (12)[1]. Since the luminosity is determined by photon flux received, which in turn is determined by the total photons emitted, we may write $L_{SN} = L_{SN,0} f^{-3/2}$, and the photon energy flux, $F_0$, as

$$F_0 = \frac{L_{SN,0}}{4\pi S_k^2(d_P)} \left(\frac{1}{(1+z)^2 f}\right) = \frac{L_{SN} f^{3/2}}{4\pi S_k^2(d_P)} \left(\frac{1}{(1+z)^2 f}\right)$$

$$= \frac{L_{SN}}{4\pi S_k^2(d_P)} \left(\frac{f^{1/2}}{(1+z)^2}\right). \quad (15)$$

Since the observed flux $F_0$ is related to the luminosity distance $d_L$,

$$F_0 = L_{SN}/4\pi d_L^2, \quad (16)$$

we get

$$d_L = S_k(d_P)(1+z)f^{-1/4}$$

---

[1] Let us assume that $\alpha$ is positive in equation (5) so that the function $f(a) = \exp(a^\alpha - 1)$ increases with the scale factor $a$. Since $N = N_0 f^{-3/2}$ (equation 12), the number of photons produced increases with decreasing $a$, hence also the luminosity. This would be perceived as if the SNIa is closer than it really is.

$$= S_k(d_P)(1+z)[\exp\{(1+z)^{-\alpha} - 1\}]^{-1/4}. \quad (17)$$

The distance modulus $\mu$ reported in the SNeIa Pantheon databases (e.g., Scolnic et al.) is by definition related to the luminosity distance $d_L$:

$$\mu \equiv 5\log_{10}\left(\frac{d_L}{1Mpc}\right) + 25 = 5\log_{10}\left(\frac{S_k(d_P)}{1Mpc}\right) + 5\log_{10}(1+z) + 25 - 1.25\log_{10}[\exp\{(1+z)^{-\alpha} - 1\}]. \quad (18)$$

Other factors may also affect the SNeIa luminosity as a standard candle. One such factor is the dependence of SNeIa luminosities on the metallicities of their host galaxies (Moreno-Raya et al. 2016). They showed the SNeIa located in high metallicities hosts are $0.14 \pm 0.10$ magnitude brighter than those in the low metallicity hosts. If we consider that the metallicities of the galaxies increase with their age, i.e., they increase with increasing scale factor $a$, and that the calibration of the SNIa standard candle luminosity is done from observations in galaxies with $a \approx 1$, we may write the magnitude decrease with $a$ as $0.14(1-a)$. This corresponds to the magnitude decreasing with increasing redshift $z$ as $0.14z/(1+z)$. Equation (18) is then modified:

$$\mu = 5\log_{10}\left(\frac{S_k(d_P)}{1Mpc}\right) + 5\log_{10}(1+z) + 25$$
$$- 1.25\log_{10}[\exp\{(1+z)^{-\alpha} - 1\}] - 0.14\left(\frac{z}{1+z}\right). \quad (19)$$

The metallicity correction to $\mu$ is perhaps an underestimate, but it will show how its inclusion impacts the cosmological parameters.

## 4 RESULTS

We will closely follow the mathematical treatment for SNeIa data fit under the VPC model presented in an earlier paper (Gupta 2020) while correcting some errors in that paper. The corrections are detailed in Appendix A. Accordingly,

$$d_P = \frac{c_0}{H_0} \int_0^z \frac{dz \exp[(1+z)^{-\alpha} - 1]}{E(z)}. \quad (20)$$

Here $H_0$ is the Hubble constant and $E(z)$ is the Peebles function derived in Appendix A, equation (A19), in terms of $a = 1/(1+z)$:

$$E(a)^2 = \exp(a^\alpha - 1)\left[\Omega_{m,0} a^{-3}\{1 + \alpha F(\alpha, a)\} + \Omega_{r,0} a^{-4} + \Omega_{\Lambda,0} \exp(a^\alpha - 1)\right] + \frac{\Omega_{k,0}}{a^2}\exp[2(a^\alpha - 1)]. \quad (21)$$

Here $\Omega_{m,0}$, $\Omega_{r,0}$, and $\Omega_{\Lambda,0}$ are the matter, radiation, and dark-energy densities at present relative to the critical energy density $\varepsilon_{c,0} \equiv 3c_0^2 H_0^2/(8\pi G_0)$, $\Omega_{k,0} = 1 - \Omega_{m,0} - \Omega_{r,0} - \Omega_{\Lambda,0}$ is the relative curvature energy density, and the function $F(\alpha, a)$ is

$$F(\alpha, a) \equiv a^3 \exp(a^\alpha - 1)\left[\int_1^a \frac{a'^{\alpha-4}}{\exp(a'^\alpha - 1)} da'\right] \quad (22)$$

The function $F$ here is not the same as used for SNeIa flux. With $d_P$ from equation (20) substituted in equation (18) or (19), we can fit the SNeIa Pantheon data (Scolnic et al. 2018), which has 1048 data points, for a range of values of $\alpha$ to see which value of $\alpha$ provides the cosmological parameters closest to those observed. We have considered a flat universe by constraining $\Omega_{k,0} = 0$, and a universe without dark energy by constraining $\Omega_{\Lambda,0} = 0$. However, these constraints do not provide as good a fit as those without such constraints. Moreover, such constraints are artificial and subjective and thus are not explored further in this paper. For fitting SNeIa data, $\Omega_{r,0} \ll \Omega_{m,0}$, and thus we have set $\Omega_{r,0} = 0$.

The results for $\Omega_{k,0} = 1 - \Omega_{m,0} - \Omega_{\Lambda,0}$, with $H_0$, $\Omega_{m,0}$ and $\Omega_{\Lambda,0}$ determined by fitting the SNeIa data for $\alpha$ ranging from 0 to 2.1 are shown in Figures 1 and 2. The Hubble parameter $h$ (not to be confused with the Planck constant) is defined as $H_0/100\ km\ s^{-1}\ Mpc^{-1}$. The $\chi^2$ values for 1045 degrees of freedom are 1033 for $\alpha < 0.9$ and 1032 for $\alpha \geq 0.9$, i.e., the quality of the data fit was essentially the same for all values of $\alpha$ we considered. It should be noted that $\alpha = 0$, yields the ΛCDM solution and, in addition, if we set $\Omega_{\Lambda,0} = 1 - \Omega_{m,0}$ (i.e. $\Omega_{k,0} = 0$) we get the flat Universe ΛCDM solution. We also see that as expected, constraining to flat Universe increased slightly the $\chi^2$ value ($\chi^2$ =1036 at $\alpha = 0$ to $\chi^2 = 1074$ at $\alpha = 2.1$ with a minimum of 1033 at $\alpha \approx 0.3$ to 0.6). In this work, we will focus mainly on the case for $\Omega_{k,0} \neq 0$ as this is most interesting for exploring alternatives to dark energy.

Since we get excellent Pantheon data fits for a range of values of the parameter $\alpha$, it is not possible to choose the value of $\alpha$ from such fits. We have to determine which value of $\alpha$ provides parameters $H_0$, $\Omega_{m,0}$ and $\Omega_{\Lambda,0}$ that correctly predict some other observed cosmological parameters, especially (a) the baryon acoustic oscillations (BAO) acoustic scale, $A_s$, and (b) The multipole moment, $l$, of the first peak in the CMB temperature anisotropy spectrum. Other parameters of interest are the deceleration parameter $q_0$ and age of the Universe $t_0$. The derivation of $A_s, l, q_0$ and $t_0$ is presented in Appendix B. We also show in this appendix how we calculate the redshift of the last scattering surface, $z_{CMB}$, in the VPC model. It turns out to be a factor $f^2$ lower than that of the standard model. It is shown in the appendix that the Planck form of the CMB spectrum and the equation of state for radiation remain unchanged in the VPC model.

Salient features of Fig. 1:

1. The Hubble parameter $h$ increases only slightly with a narrow band of 95% confidence bound.
2. The mass density parameter $\Omega_{m,0}$ increases slightly from 0.32 to 0.33, going from $\alpha = 0$ to 0.6 and then decrease to 0.27, i.e., it is not significantly different from the Planck (Planck Collaboration I 2020) value of $\Omega_{m,0} = 0.31$. It also has a narrow band of 95% confidence bound.
3. The dark-energy density $\Omega_{\Lambda,0}$ declines rapidly with increasing $\alpha$ whereas the curvature energy density $\Omega_{k,0}$ increases rapidly. They both have pretty wide bands of 95% confidence bound.
4. In the VPC model, we conclude that the increase in the curvature energy density compensates for the decline in dark-energy density as we increase the value of $\alpha$, i.e., we increase the strength of the constants' variation.
5. The case $\alpha = 0$ corresponds to a ΛCDM universe *without* the flat universe constraint. While $\Omega_{m,0} = 0.3168 \pm 0.0709$ is reasonably close to the accepted value, $\Omega_{\Lambda,0} = 0.9126 \pm 0.1224$ is significantly higher, meaning that $\Omega_{k,0} = -0.2294 \pm 0.1933$. What is not shown in the figure, but is important to mention, is that *with* the flat universe constraint, we get $\Omega_{m,0} = 0.2845 \pm 0.0244$ and $\Omega_{\Lambda,0} = 0.7155 \pm 0.0244$, the same as one would expect.

Key features of Fig. 2:

1. The variation of $A_s$ and $l$ with $\alpha$ relative to their Planck values ($A_s = 147.57$ Mpc and $l = 220.6$) show that $A_s$ and $l$ both have their Planck values between $\alpha = 1.5$ and 2.0.
2. The contour plot of $A_s$ and $l$, corresponding to $\alpha = 1.8$, with $\Omega_{m,0}$ and $\Omega_{\Lambda,0}$ varying over their 95% confidence bound, yields $\Omega_{m,0} =$ 0.2625 and $\Omega_{\Lambda,0} = 0.2495$ at $A_s = 147.57$ Mpc and $l = 220.6$. We have chosen $\alpha = 1.8$ for the contour plot based on an earlier work that showed this value analytically (Gupta 2018).
3. The age of the Universe, $t_0 = 14.22$ Gyr at $\alpha = 0$ is about the same as the Planck value (13.787 Gyr) but increases steadily to 19.73 Gyr at $\alpha = 2.1$, with 19.12 Gyr at $\alpha = 1.8$.
4. The deceleration parameter, $q_0 = -0.75$ at $\alpha = 0$ (not constrained to a flat universe) is more significant than its Planck value ($-0.5341$) and steadily approaches $-2.02$ at $\alpha = 2.1$, with $-1.84$ at $\alpha = 1.8$.
5. The redshift $z_{CMB}$, corresponding to the surface of last scattering, decreases rapidly with increasing $\alpha$ initially but then plateaus to 146.5 above $\alpha \approx 1.2$.
6. All constants, $c, G, h,$ and $k_B$ decrease with decreasing scale factor $a$ but approach their respective fixed values as $a \to 0$.

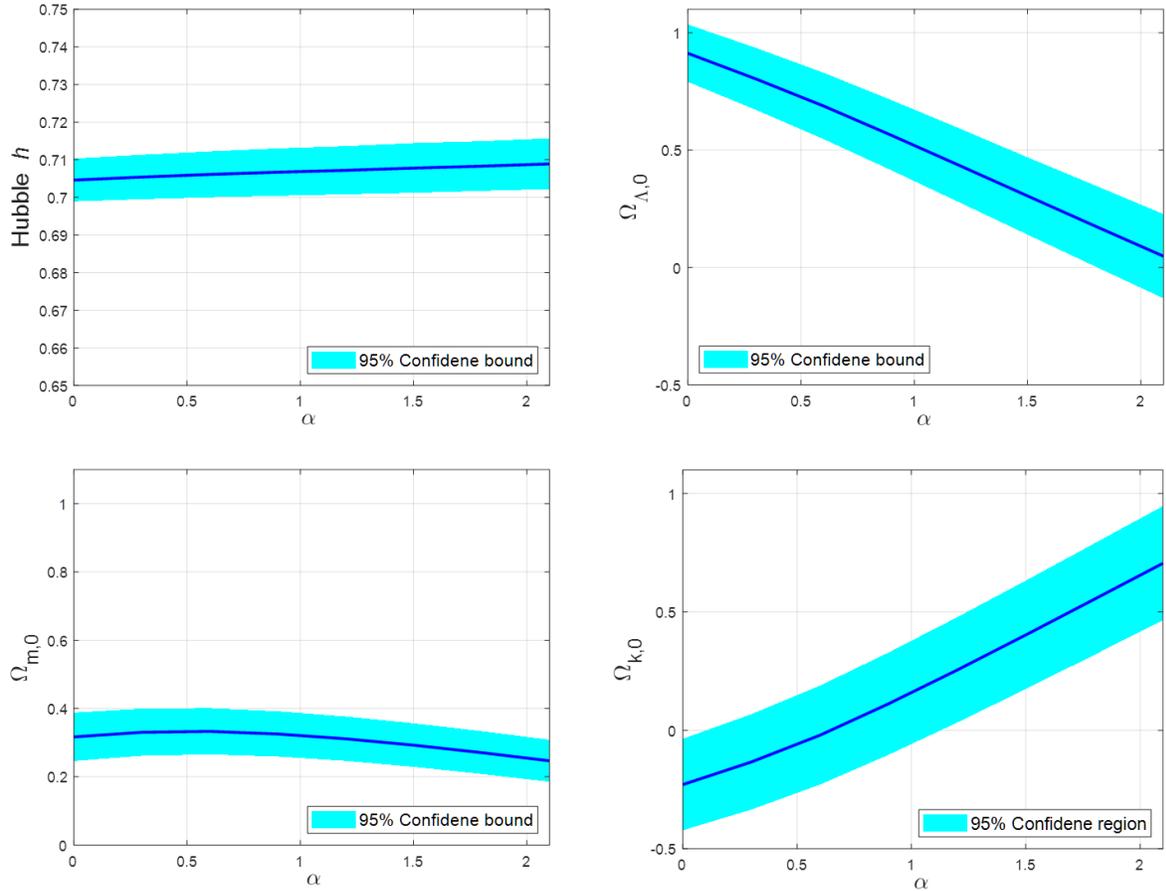

Fig. 1: Key cosmological parameters, Hubble $h$, $\Omega_{\Lambda,0}$, $\Omega_{m,0}$, and $\Omega_{k,0}$ in the VPC model. The parameters are determined by fitting Pantheon SNeIa data. As discussed in the text, the SNeIa luminosities are corrected for the Chandrasekhar mass effect and metallicities. These parameters were evaluated for increasing values of the parameter $\alpha$ used for defining $f(\alpha)$. While there is a relatively small dependence of $h$ and $\Omega_{m,0}$ on $\alpha$, $\Omega_{\Lambda,0}$ and $\Omega_{k,0}$ depend strongly on $\alpha$; $\Omega_{\Lambda,0}$ decreases whereas $\Omega_{k,0}$ increases with $\alpha$. This means the curvature energy density compensates for the dark-energy density in the VPC model. The 95% confidence band ($2\sigma$) is significantly narrower for $\Omega_{m,0}$ than for $\Omega_{\Lambda,0}$ and $\Omega_{k,0}$.

Let us see how the luminosity of an SNeIa is affected with the scale factor $a$ and the VPC strength parameter $\alpha$ using the relations $L_{SN} = L_{SN,0} f^{-3/2}$. Since $f \equiv \exp(a^\alpha - 1) \to 1/e = 0.368$ as $a \to 0$, we see that $f^{-3/2} \to$ 4.48 as $a \to 0$ for all positive non-zero values of $\alpha$. As shown in Fig. 3, the relative luminosity of SNeIa, $L_{SN}/L_{SN,0}$, approaches the limiting value 4.48 more and more rapidly with increasing $\alpha$. Considering that the luminosity distance is proportional to the square root of the luminosity, we could say the luminosity distance determined from SNeIa luminosity could be higher by up to a factor of 2.12 in the VPC model.

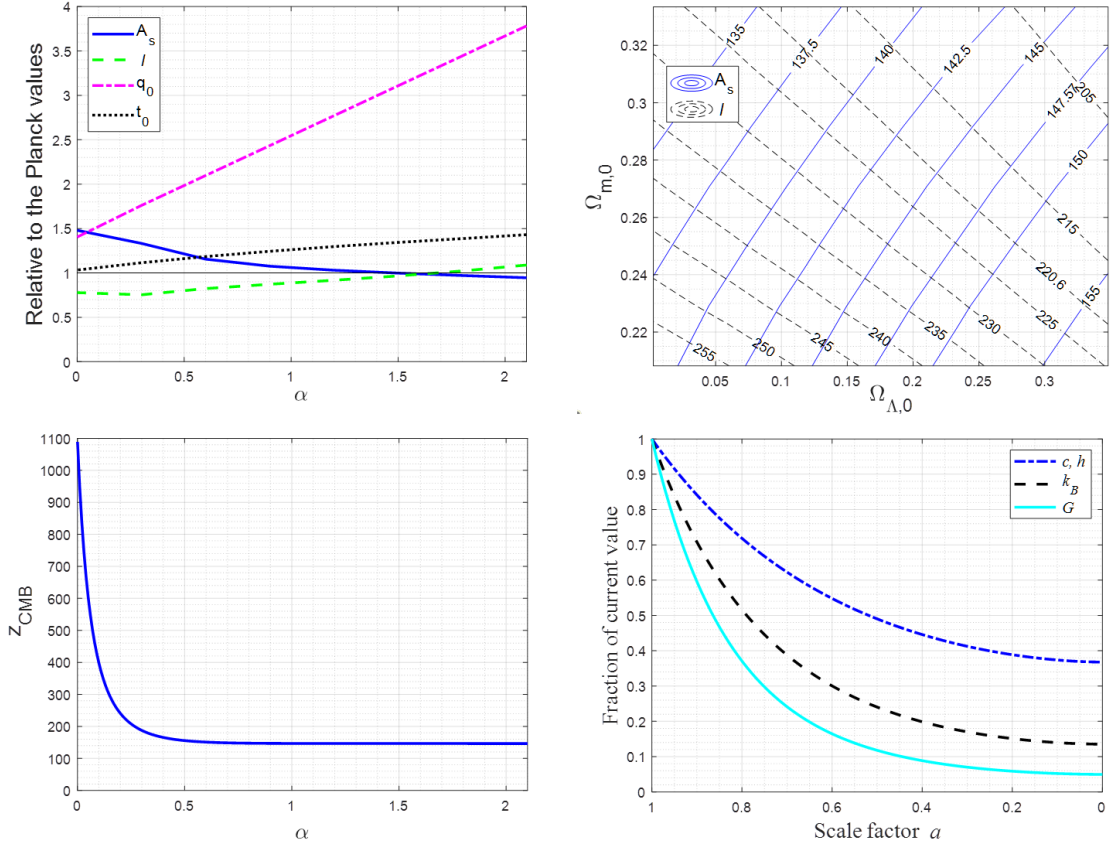

Fig. 2: Variation of BAO acoustic scale $A_s$, CMB multipole moment $l$, deceleration parameter $q_0$, age of the Universe $t_0$, and the redshift of the last scattering surface $z_{CMB}$, against coupling constants' strength parameter $\alpha$, and variation of the coupling constants against the scale factor $a$. *Top-left*: The dependence on $\alpha$ of $A_s, l, q_0,$ and $t_0$, relative to their Planck values ($A_s = 147.57$ Mpc, $l = 220.6, q_0 = -0.5341,$ and $t_0 = 13.787$ Gyr). The most dramatic change is seen in $q_0$. Parameters $A_s$ and $l$ acquire their Planck values between $\alpha = 1.5$ and $2.0$. *Top-right*: Contour plot of $A_s$ and $l$ in the $(\Omega_{m,0}, \Omega_{\Lambda,0})$ space at $\alpha = 1.8$. It is used to determine, corresponding to the Planck values, the values of $\Omega_{m,0}$ and $\Omega_{\Lambda,0}$ shown with variance in Fig. 1. *Bottom-left*: The variation of $z_{CMB}$ with $\alpha$. Initially, it decreases rapidly with increasing $\alpha$ but then plateaus to 146.5 for $\alpha > 1.2$. *Bottom-right*: Variation at $\alpha = 1.8$ of the speed of light $c$, Planck constant $h$, the Boltzmann constant $k_B$, and the gravitational constant $G$, with the scale factor $a$. They all approach constant values as $a \to 0$.

## 5 DISCUSSION

Wright and Li (2018) considered the variation of gravitational constant given by $G = G_0(1+z)^{-1/n}$ with $n = 4$, i.e., $G = G_0(1+z)^{-1/4}$, and showed that the SNeIa data could be fitted equally well with $(\Omega_{m,0}, \Omega_{\Lambda,0}) = (0.62, 0.38)$ with said $G$ variation, and with $(\Omega_{m,0}, \Omega_{\Lambda,0}) = (0.3, 0.7)$ without any $G$ variation using the standard model, assuming in both cases a flat universe. They did not attempt to change or optimize the exponent of $(1+z)$, i.e., the value of $n$. Wright (2021) contended that without some strong priors on $(\Omega_{m,0}, \Omega_{\Lambda,0})$ it is not possible to determine the optimum value of $n$, presumably because different values of $n$ would yield different $(\Omega_{m,0}, \Omega_{\Lambda,0})$ values upon fitting the SNeIa data. This means that for any cosmological model, the SNeIa data fit is *an essential but not sufficient* condition for establishing the model's validity and the parameters the model thus determines. Additionally, they did not consider the variation of $c$ and $h$.

We see essentially the same problem in our approach. We cannot determine a unique value of $\alpha$ by fitting the SNeIa data as all plausible values of $\alpha$ we have tried, from $\alpha = 0$ to $\alpha = 2.1$, provide equally good data fit. We then took the parameters $H_0$, $\Omega_{m,0}$ and $\Omega_{\Lambda,0}$ for each $\alpha$ value, and calculated corresponding BAO acoustic scale $A_s$ and the multipole moment $l$ for the first CMB anisotropy peak. Since $A_s$ and $l$ are observable quantities, they were used to determine the $\alpha$ value that yields observed $A_s$ and $l$ values. The value lies (Fig. 2) between $\alpha = 1.5$ and 2.0. We can therefore take $\alpha = 1.8$, the value that was determined analytically by relaxing the assumption of an adiabatically expanding Universe (Gupta 2018). The 95% confidence bounds of $(\Omega_{m,0}, \Omega_{\Lambda,0})$ for $1.5 < \alpha < 2.0$ have significant overlap (Fig. 1). Thus, $\alpha = 1.8$ could yield acceptable $(\Omega_{m,0}, \Omega_{\Lambda,0})$ pair values, as well as the correct $A_s$ and $l$ values. We, therefore, plotted the contours of $A_s$ and $l$ within the 95% confidence bounds of $(\Omega_{m,0}, \Omega_{\Lambda,0})$ at $\alpha = 1.8$ and determined $(\Omega_{m,0}, \Omega_{\Lambda,0}) = (0.2625, 0.2495)$ corresponding to Planck's $(A_s, l) = (147.57, 220.6)$ for the VPC model (Fig. 2). We could therefore safely state that the $\alpha = 1.8$ VPC model giving $(\Omega_{m,0}, \Omega_{\Lambda,0}) = (0.2708_{0.2082}^{0.3334}, 0.1754_{0.0027}^{0.3481})$ is an acceptable cosmological model. Interestingly, the matter energy density $\Omega_{m,0}$ does not depend significantly on the variability of the constants' strength parameter $\alpha$, but the dark energy density $\Omega_{\Lambda,0}$ does; it decreases with $\alpha$ whereas the curvature energy density, $\Omega_{k,0}$, increases. The insignificant variation of $\Omega_{m,0}$ with $\alpha$ is vital since $\Omega_{m,0}$ can be constrained directly from observations, such as from galactic cluster abundances (e.g., Abdullah et al. 2020) at $\Omega_{m,0} = 0.310_{0.283}^{0.333}$. But this also means that $\Omega_{m,0}$ cannot constrain $\alpha$. This is the reason we had to constrain $\alpha$ using $(A_s, l)$ pair values.

The deceleration parameter $q_0$ is a model-dependent parameter and could be significantly different depending on the model, data used, and corrections applied to the data. Using local observations, cosmological principle, and flat Universe prior, Camarena and Marra (2020) determined $q_0 = -1.08_{-1.37}^{-0.79}$, which disagrees with Planck's value, $q_0 = -0.5341$, at $1.9\sigma$ level. The VPC value, $q_0 = -1.84$ at $\alpha = 1.8$, we have determined, disagrees with Planck's value even more significantly. However, none of these values is supported from observations, and thus a model cannot be rejected or accepted based on the value of the deceleration parameter it generates.

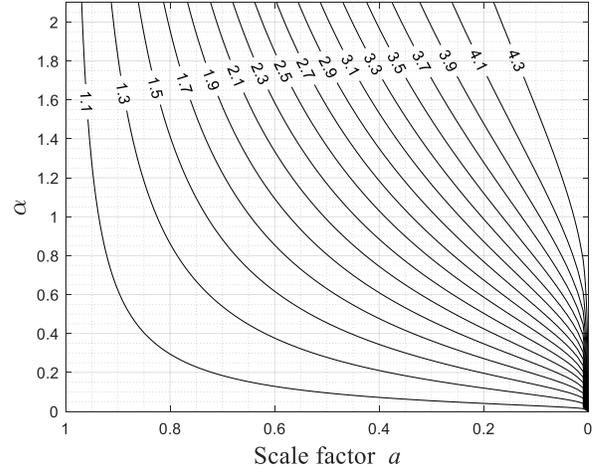

Fig. 3: The contour plot of SNeIa relative luminosity $(L_{SN}/L_{SN,0})$ in $(a, \alpha)$ space of the VPC model.

Coming now to the age of the Universe, $t_0$, we see it to be 39% higher for the VPC model with $\alpha = 1.8$ ($t_0 = 19.12$ Gyr) than that determined by Planck ($t_0 = 13.787$ Gyr). However, some of the oldest stars in our Galaxy, which are near enough to measure their distance by parallax method with a high degree of confidence, have ages greater than 13.787 Gyr. The star HD140283 is only about 58 parsecs away from us with an age of $14.46 \pm 0.31$ Gyr (Bond et al. 2013, VandenBerg et al. 2014). In fact, the age of this star was initially determined to be $\sim$16 Gyr but was revised downward by 'correcting' some model parameters and measurements. And HD140283 does not even belong to a globular star cluster; globular clusters are known to be populated with the oldest stars. The nearest globular cluster is $\sim$1700 parsec away, and thus parallax method cannot be used for determining stellar distances. For many astrophysicists, a higher than the Planck's age of the Universe will be a welcome news as their research findings will not have to be constrained by the 13.787 Gyr age of the Universe. The stellar age determination depends on the stellar evolution model and parameters used in the model. Therefore, one may discard a model that yields any stellar age exceeding the generally accepted Universe age. Moreover, the models are based on non-evolutionary physical constants and on applying local energy conservation laws to the problems involving evolution at a cosmological time scale where energy is not conserved (e.g., Peebles 1993, Baryshev 2008). A salient point to note is that the age determined in the VPC model involves an increasing speed of light and thus should be expected to yield higher Universe age.

Now, relations among physical constants in the VPC model, given by equation (5), maybe written as $\dot{c}/c = \alpha a^\alpha (\dot{a}/a) = \alpha a^\alpha H$; $\dot{G}/G = 3\alpha a^\alpha H$, and $\dot{h}/h = \alpha a^\alpha H$, i.e. $\dot{G}/G = 3\dot{c}/c = 3\dot{h}/h$. Thus, in the VPC model, $c$, $G$, and $h$ vary concurrently, and their variations are related. One could therefore conclude that the present constraint ($a = 1$) on $(\dot{G}/G)_0 = 5.4 H_0$ yields an excellent fit to the Pantheon SNeIa data when using the VPC model. Indeed the bounds on $(\dot{G}/G)_0$ depend on the bounds on $H_0$ which in turn depends on the cosmological model used. Taking $H_0 = 70.83(\pm 0.66)$ km $s^{-1}$ Mp$c^{-1}$ = 7.23 $(\pm 0.07) \times 10^{-11}\ yr^{-1}$, we get $(\dot{G}/G)_0 = 3.90(\pm 0.04) \times 10^{-10}\ yr^{-1}$, against $(\dot{G}/G)_0 \leq 10^{-11}\ yr^{-1}$ obtained by Gaztañaga et al. (2001). Wright and Li (2018) determined $G = G_0(1+z)^{-1/4} \equiv G_0 a^{1/4}$. This translates into $(\dot{G}/G)_0 = 0.25 H_0 = 1.8 \times 10^{-11}\ yr^{-1}$ (and has the same sign as of this paper). Thus, our $(\dot{G}/G)_0$ constraint at present is an order of magnitude higher than theirs as they treat only $G$ to be varying whereas $c$ and $h$ are kept invariant. We also see that the constraints on $c$ and $h$ are both one-third of the constraint on $G$, i.e. $(\dot{c}/c)_0 = (\dot{h}/h)_0 = 1.30(\pm 0.01) \times 10^{-10}\ yr^{-1}$ at present.

If we compare our value of $(\dot{G}/G)_0$ with that determined from orbital timing measurements, e.g., Hofmann & Müller 2018: $(\dot{G}/G)_0 = (7.1 \pm 7.6) \times 10^{-14}\ yr^{-1}$, then the latter is several orders of magnitude lower since the speed of light variation is not considered in such works. When the speed of light is allowed to vary the same way as in this paper, then it can be shown that the orbital timing constraints determined are on $(\dot{G}/G - 3\dot{c}/c)_0$ and not on $(\dot{G}/G)_0$ (Gupta 2021a).

One additional point we wish to address is that the variation of the speed light tantamounts to the breaking of Lorentz symmetry. Nagel et al. (2015) have tested Lorentz symmetry in electrodynamics using Michelson-Morley interferometer to a level of $\Delta c/c \sim 10^{-18}$. More recently, Kostelecky et al. (2016) did the same by analyzing LIGO data to a level of $\Delta c/c \sim 10^{-21}$. Such works establish the spatial invariance of the speed of light rather than the temporal invariance, that too in the local frame rather than at a cosmological scale.

Suggestions have recently been made that the temporal variation of $c$ at cosmological time scale is possible to test with a combination of strongly lensed and unlensed SNeIa observations (Cao et al. 2018, 2020). Accordingly, the measurement of time delay difference in the peaking of the SNeIa luminosity between two gravitationally lensed images of the SNeIa should enable one to determine if $c$ was different from its current value at the time the lensing galaxy produced the images. However, since the geometrical time delay and the Shapiro time delay both scale as $G/c^3$ (e.g., Dodelson 2017) it can be shown that this method of constraining $c$ is valid only when the possible variation of the gravitational constant $G$ is ignored: the $c$ variation is imperceptible when $G$ varies as $c^3$ (Gupta 2021c).

It should be mentioned that the same variation of physical constants used in this paper, when applied to the big-bang nucleosynthesis, is able to resolve the so-called 'lithium problem' (Gupta 2021b), further increasing our confidence in the VPC model. However, unless we can design an experiment or an observation that can directly determine the cosmological variation of even one of the physical constants, the skepticism in their variation will continue to prevail.

# 6 CONCLUSION

The conclusions of this paper may be summarized as follows:

1. The supernova type 1a (SNIa) standard candle luminosity evolves as $\exp[1.5(1 - a^\alpha)]$ where $a$ is the scale factor, and $\alpha$ is the strength parameter of the VPC mode. This means that the luminosity of distant type 1a supernovae could be up to $\approx 4$ times higher in the VPC model than in the standard model. This affects the luminosity distance by a factor of up to $\approx 2$.
2. All plausible values of $\alpha$ provide equally good fits to the Pantheon SNeIa data, and yield about the same matter density, $\Omega_{m,0} \approx 0.3$, while the dark energy density and curvature energy densities trade their values with each other. Thus, a correct $\alpha$ value cannot be determined just by fitting the SNeIa data.
3. The observed and Planck verified BAO acoustic scale factor, $A_s = 147.57$, and the multipole moment value of the first CMB temperature anisotropy peak, $l = 220.6$, constrain $\alpha = 1.8$.
4. Variations of the constants are given by $(\dot{G}/G)_0 = 3.90(\pm 0.04) \times 10^{-10}\ yr^{-1}$ and $(\dot{c}/c)_0 = (\dot{h}/h)_0 = 1.30(\pm 0.01) \times 10^{-10}\ yr^{-1}$ at present. These variations are valid only when $G$, $c$, and $h$ are permitted to vary concurrently rather than individually.

**ACKNOWLEDGEMENT**

A grant from Macronix Research Corporation partially supported this research. The author wishes to thank Dr. Bill Wright for his email to communicate the limitation of his work that we have referred to in this paper. He is grateful to Dr. Santanu Das and Dr. Rodrigo Cuzinatto for innumerable discussions on Zoom and in-person, and many email communications.

**DATA AVAILABILITY**

The data used in this paper is available readily from the author.

**APPENDIX A: GENERAL CONSTRAINT AND PROPER DISTANCE**

We will closely follow an earlier publication (Gupta 2020) to minimize the mathematics already presented in that paper.

The general constraint is written as (Costa et al. 2019)

$$\left[\left(\frac{\dot{G}}{G} - 4\frac{\dot{c}}{c}\right)\frac{8\pi G}{c^4}\varepsilon + \dot{\Lambda}\right] = 0 \tag{A1}$$

Here $G$ is the gravitational constant, $c$ is the speed of light, $\varepsilon$ is the energy density, and $\Lambda$ is the cosmological constant. All the constants are allowed to vary with the scale factor $a$. Taking $\dot{G}/G = 3\dot{c}/c$, we may write:

$$\frac{\dot{c}}{c}\frac{8\pi G}{c^4\Lambda}\varepsilon = \frac{\dot{\Lambda}}{\Lambda} \Rightarrow \frac{c^4\Lambda}{8\pi G} \equiv \varepsilon_\Lambda = \frac{\dot{c}}{c}\frac{\Lambda}{\dot{\Lambda}}\varepsilon \tag{A2}$$

Let us define $c(a) \equiv c_0 f(a)$ and $\Lambda(a) \equiv \Lambda_0 g(a)$, then $\dot{c}/c = \dot{a}f'(a)/f(a)$; $\dot{\Lambda}/\Lambda = \dot{a}g'(a)/g(a)$; prime, ', indicates the $a$ derivative. [Due to the general constraint equation (A1), function $f(a)$ and $g(a)$ are not independent, and one must find the relationship between the two. This fact was not considered in Gupta (2020).] We may now write

$$\varepsilon_\Lambda = \frac{c^4\Lambda}{8\pi G} = \frac{\dot{a}f'(a)}{f(a)}\frac{g(a)}{\dot{a}g'(a)}\varepsilon \Rightarrow \frac{c_0^4\Lambda_0}{8\pi G_0}f(a)g(a) = \frac{f'(a)}{f(a)}\frac{g(a)}{g'(a)}(\varepsilon_{m,0}a^{-3} + \varepsilon_{r,0}a^{-4}). \tag{A3}$$

Here $\varepsilon_m$ is the matter energy density and $\varepsilon_r$ is the radiation energy density. The subscript 0 refers to the parameter value at the current time, i.e., for $a = 1$. Rearranging we get

$$g'(a) = \frac{\varepsilon}{\varepsilon_{\Lambda,0}}\frac{f'(a)}{f(a)^2} = \frac{1}{\varepsilon_{\Lambda,0}}\frac{f'(a)}{f(a)^2}(\varepsilon_{m,0}a^{-3} + \varepsilon_{r,0}a^{-4}). \tag{A4}$$

If we write $f(a) = a^\alpha$ and $g(a) = a^\beta$ in the power-law form, we have $f'(a) = \alpha a^{\alpha-1}$ and $g'(a) = \beta a^{\beta-1}$. Then,

$$\beta a^{\beta-1} = \frac{1}{\varepsilon_{\Lambda,0}}\frac{\alpha a^{\alpha-1}}{a^{2\alpha}}(\varepsilon_{m,0}a^{-3} + \varepsilon_{r,0}a^{-4}). \tag{A5}$$

This equality is satisfied only when $\alpha + \beta = -3$ in matter-dominated Universe and $\alpha + \beta = -4$ in radiation dominated Universe. However, the power-law forms lead to singularities as $a \to 0$. Our approach, therefore, has been to use an alternative form $f(a) = \exp(a^\alpha - 1)$, then $f'(a) = \alpha a^{\alpha-1}f(a)$. Equation (A4) is then

$$g'(a) = \frac{\varepsilon}{\varepsilon_{\Lambda,0}}\frac{\alpha a^{\alpha-1}}{\exp(a^\alpha-1)} = \frac{\varepsilon_{m,0}}{\varepsilon_{\Lambda,0}}\frac{\alpha a^{\alpha-4}}{\exp(a^\alpha-1)}\left(1 + \frac{\varepsilon_{r,0}}{\varepsilon_{m,0}}a^{-1}\right). \tag{A6}$$

This leads to

$$g_\alpha(a) = 1 + \frac{\varepsilon_{m,0}\alpha}{\varepsilon_{\Lambda,0}}\int_1^a \frac{a^{\alpha-4}}{\exp(a^\alpha-1)}\left(1 + \frac{\varepsilon_{r,0}}{\varepsilon_{m,0}}a^{-1}\right)da, \tag{A7}$$

since $g_\alpha(a=1) = 1$. We may now write from equation (A3)

$$\varepsilon_\Lambda = \frac{f'(a)}{f(a)}\frac{g(a)}{g'(a)}\varepsilon = \alpha a^{\alpha-1}\left[1 + \frac{\varepsilon_{m,0}\alpha}{\varepsilon_{\Lambda,0}}\int_1^a \frac{a'^{\alpha-4}}{\exp(a'^\alpha-1)}\left(1 + \frac{\varepsilon_{r,0}}{\varepsilon_{m,0}}a'^{-1}\right)da'\right]\left[\frac{\varepsilon}{\varepsilon_{\Lambda,0}}\frac{\alpha a^{\alpha-1}}{\exp(a^\alpha-1)}\right]^{-1}\varepsilon$$

$$= \varepsilon_{\Lambda,0}\exp(a^\alpha - 1)\left[1 + \frac{\varepsilon_{m,0}\alpha}{\varepsilon_{\Lambda,0}}\int_1^a \frac{a'^{\alpha-4}}{\exp(a'^\alpha-1)}\left(1 + \frac{\varepsilon_{r,0}}{\varepsilon_{m,0}}a'^{-1}\right)da'\right]. \tag{A8}$$

Equation (14) in Gupta (2020) now becomes

$$H^2 = \frac{8\pi G}{3c^2}\left(\varepsilon + \frac{\Lambda c^4}{8\pi G}\right) - \frac{kc^2}{a^2} = \frac{8\pi G}{3c^2}(\varepsilon + \varepsilon_\Lambda) - \frac{kc^2}{a^2}$$

$$= \frac{8\pi G}{3c^2}\left(\varepsilon + \varepsilon_{\Lambda,0}\exp(a^\alpha - 1)\left[1 + \frac{\varepsilon_{m,0}\alpha}{\varepsilon_{\Lambda,0}}\int_1^a \frac{a'^{\alpha-4}}{\exp(a'^\alpha-1)}\left(1 + \frac{\varepsilon_{r,0}}{\varepsilon_{m,0}}a'^{-1}\right)da'\right]\right) - \frac{kc^2}{a^2}, \text{ or} \tag{A9}$$

$$\frac{H^2}{H_0^2} = \frac{8\pi G}{3c^2 H_0^2}\left(\varepsilon + \varepsilon_{\Lambda,0}\exp(a^\alpha - 1)\left[1 + \frac{\varepsilon_{m,0}\alpha}{\varepsilon_{\Lambda,0}}\int_1^a \frac{a'^{\alpha-4}}{\exp(a'^\alpha-1)}\left(1 + \frac{\varepsilon_{r,0}}{\varepsilon_{m,0}}a'^{-1}\right)da'\right]\right) - \frac{kc^2}{a^2 H_0^2}, \text{ or at } a = 1, \text{ or} \tag{A10}$$

$$1 = \frac{8\pi G_0}{3c_0^2 H_0^2}[\varepsilon_0 + \varepsilon_{\Lambda,0}] - \frac{kc_0^2}{H_0^2} = \frac{1}{\varepsilon_{c,0}}[\varepsilon_0 + \varepsilon_{\Lambda,0}] - \frac{kc_0^2}{H_0^2} \equiv \Omega_{m,0} + \Omega_{r,0} + \Omega_{\Lambda,0} - \frac{kc_0^2}{H_0^2}. \tag{A11}$$

Here critical density is $\varepsilon_{c,0} \equiv \frac{3c_0^2 H_0^2}{8\pi G_0}$. By defining $\Omega_0 \equiv \Omega_{m,0} + \Omega_{r,0} + \Omega_{\Lambda,0}$, we may write

$$\Omega_{k,0} \equiv -\frac{kc_0^2}{H_0^2} = 1 - \Omega_0. \tag{A12}$$

Therefore,

$$\frac{H^2}{H_0^2} = \exp(a^\alpha - 1)\left[\Omega_{m,0}a^{-3} + \Omega_{r,0}a^{-4} + \Omega_{\Lambda,0}\exp(a^\alpha - 1)\left[1 + \frac{\varepsilon_{m,0}\alpha}{\varepsilon_{\Lambda,0}}\int_1^a \frac{a'^{\alpha-4}}{\exp(a'^\alpha-1)}\left(1 + \frac{\varepsilon_{r,0}}{\varepsilon_{m,0}}a'^{-1}\right)da'\right]\right]$$

$$+ \frac{1 - \Omega_{m,0} - \Omega_{r,0} - \Omega_{\Lambda,0}}{a^2}\exp[2(a^\alpha - 1)], \text{ or} \tag{A13}$$

$$\frac{H^2}{H_0^2} = \exp(a^\alpha - 1)\left[\Omega_{m,0}a^{-3} + \Omega_{r,0}a^{-4} + \Omega_{\Lambda,0}\exp(a^\alpha - 1)\left[1 + \frac{\Omega_{m,0}\alpha}{\Omega_{\Lambda,0}}\int_1^a \frac{a'^{\alpha-4}}{\exp(a'^\alpha-1)}\left(1 + \frac{\Omega_{r,0}}{\Omega_{m,0}}a'^{-1}\right)da'\right]\right]$$

$$+ \frac{1 - \Omega_{m,0} - \Omega_{r,0} - \Omega_{\Lambda,0}}{a^2}\exp[2(a^\alpha - 1)], \text{ or} \tag{A14}$$

$$\frac{H^2}{H_0^2} = \exp(a^\alpha - 1)\left[\Omega_{m,0}a^{-3} + \Omega_{r,0}a^{-4} + \left[\Omega_{\Lambda,0}\exp(a^\alpha - 1) + \Omega_{m,0}\exp(a^\alpha - 1)\alpha\int_1^a \frac{a'^{\alpha-4}}{\exp(a'^\alpha-1)} \times \left(1 + \frac{\Omega_{r,0}}{\Omega_{m,0}}a'^{-1}\right)da'\right]\right]$$
$$+ \frac{1 - \Omega_{m,0} - \Omega_{r,0} - \Omega_{\Lambda,0}}{a^2}\exp[2(a^\alpha - 1)], \text{ or} \tag{A15}$$

$$\frac{H^2}{H_0^2} = \exp(a^\alpha - 1)\left[\Omega_{m,0}a^{-3}\left\{1 + \alpha\exp(a^\alpha - 1)a^3\int_1^a \frac{a'^{\alpha-4}}{\exp(a'^\alpha-1)}\left(1 + \frac{\Omega_{r,0}}{\Omega_{m,0}}a'^{-1}\right)da'\right\} + \Omega_{r,0}a^{-4} + \Omega_{\Lambda,0}\exp(a^\alpha - 1)\right]$$
$$+ \frac{1 - \Omega_{m,0} - \Omega_{r,0} - \Omega_{\Lambda,0}}{a^2}\exp[2(a^\alpha - 1)] \tag{A16}$$

If we define a function $F$ as

$$F(\alpha, a) \equiv a^3 \exp(a^\alpha - 1)\left[\int_1^a \frac{a'^{(\alpha-4)}}{\exp(a'^\alpha-1)}\left(1 + \frac{\Omega_{r,0}}{\Omega_{m,0}}a'^{-1}\right)da'\right], \tag{A17}$$

we may write equation (A16) as

$$\frac{H^2}{H_0^2} = \exp(a^\alpha - 1)\left[\Omega_{m,0}a^{-3}\{1 + \alpha F(\alpha, a)\} + \Omega_{r,0}a^{-4} + \Omega_{\Lambda,0}\exp(a^\alpha - 1)\right] + \frac{1 - \Omega_{m,0} - \Omega_{r,0} - \Omega_{\Lambda,0}}{a^2}\exp[2(a^\alpha - 1)] \tag{A18}$$

$$= E(a)^2 \to E(z)^2 \text{ by substituting } a = 1/(1+z). \tag{A19}$$

When $\alpha = 0$, this expression assumes the form for the $\Lambda$CDM model:

$$\frac{H^2}{H_0^2} = \left(\Omega_{m,0}a^{-3} + \Omega_{r,0}a^{-4} + \Omega_{\Lambda,0}\right) - \frac{1-\Omega_{m,0}-\Omega_{r,0}-\Omega_{\Lambda,0}}{a^2}. \tag{A20}$$

(This should be expected and provides an essential check on the correctness of the above derivation.) Let us consider the behaviour of the function $F(\alpha, a)$, equation (A17). Since $\Omega_{r,0} \ll \Omega_{m,0}$, the second term, $\frac{\Omega_{r,0}}{\Omega_{m,0}}a'^{-1}$, in the bracket can be neglected when $a$ is not very small since $\Omega_{r,0} \ll \Omega_{m,0}$. When $a \ll 1$, this term can indeed be significant. However, $F(\alpha, a) \ll 1$ at $a \ll 1$ due to the factor $a^3 \exp(a^\alpha - 1)$, and could be neglected in equation (A18). The net result is that $\frac{\Omega_{r,0}}{\Omega_{m,0}}a'^{-1}$ can be safely ignored in equation (A17), i.e., $F$ maybe considered independent of $\Omega_{m,0}$ and $\Omega_{r,0}$. A numerical check to see how $F$ behaves at $a \ll 1$ has confirmed this premise. We may, therefore, write

$$F(\alpha, a) \equiv a^3 \exp(a^\alpha - 1) \left[\int_1^a \frac{a'^{\alpha-4}}{\exp(a'^\alpha-1)} da'\right]. \tag{A21}$$

The behaviour of the function $F$ is shown graphically in Fig. A1 for $\alpha = 1.8$. One can thus fit it with a polynomial and the polynomial used to represent the function to avoid evaluating time-consuming double integrals in equations (20) to (22) in Section 4.

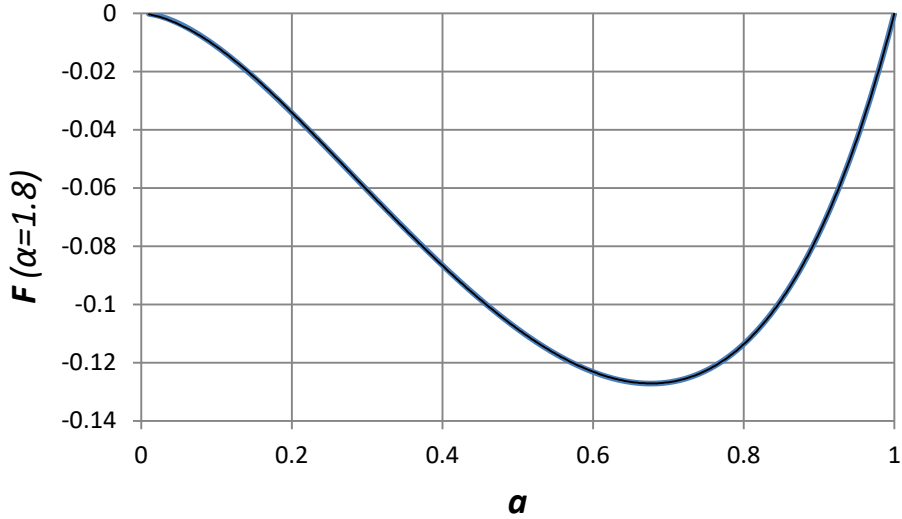

Fig. A1. The behaviour of the function $F$ against the scale factor $a$ for $\alpha = 1.8$.

We will now determine the proper distance $d_P$ between an observer and a source. We may write the FLRW (Friedmann–Lemaître–Robertson–Walker) metric in spherical spatial coordinates as (Ryden 2017)

$$ds^2 = -c^2 dt^2 + a(t)^2 [dr^2 + S_k(r)^2 (d\theta^2 + \sin^2\theta d\phi^2)]. \tag{A22}$$

Here $S_k(r) = R\sin(r/R)$ for $k = +1$ (closed Universe); $S_k(r) = r$ for $k = 0$ (flat Universe), $S_k(r) = R\sinh(r/R)$ for $k = -1$ (open Universe), where $R$ is the parameter related to the curvature. The proper distance $d_P$ is determined at fixed time by following a spatial geodesic at constant $\theta$ and $\phi$. Then

$$ds = a(t)dr \Rightarrow d_P(t) = a(t)\int_0^r dr = a(t)r. \tag{A23}$$

We could determine $r$ following a null geodesic from the time $t$ a photon is emitted by the source to the time $t_0$ it is detected by the observer with $ds = 0$ in equation (A22) at constant $\theta$ and $\phi$:

$$c^2 dt^2 = a(t)^2 dr^2 \Rightarrow \frac{cdt}{a(t)} = dr \Rightarrow r = \int_0^r dr = \int_t^{t_0} \frac{cdt}{a(t)} \Rightarrow d_P = a(t_0) \int_t^{t_0} \frac{cdt}{a(t)}. \qquad (A24)$$

Now $dt = da.dt/da = da/\dot{a} = da/a\dot{a}/a = da/aH$, and $a = 1/(1+z)$, $da = -dz/(1+z)^2 = -a^2 dz$. Therefore

$$dt = -\frac{adz}{H} = -\frac{adz}{\frac{H_0 H}{H_0}} = -\frac{adz}{H_0 E(a)}, \text{ and} \qquad (A25)$$

$$d_P = \frac{1}{H_0} \int_0^z \frac{cdz}{E(z)} = \frac{c_0}{H_0} \int_0^z \frac{\exp[((1+z)^{-\alpha} - 1)]dz}{E(z)}. \qquad (A26)$$

Here $E(z)$ is given by equation (A18, A19).

**APPENDIX B: OTHER COSMOLOGICAL PARAMETERS**

*Multipole moment of the first peak in the CMB temperature anisotropy spectrum*: The first thing is to determine the sound horizon distance $d_s(t_{ls})$, the distance sound travels at the speed $c_s(t)$ in photon-baryon fluid from the big-bang until the time such plasma disappeared due to the formation of the atoms, i.e., the time of the last scattering $t_{ls}$. Following equation (A24), we may write (Durrer 2008)

$$d_{sh}(t_{ls}) = a(t_{ls}) \int_0^{t_{ls}} \frac{c_s(t)dt}{a(t)}. \qquad (B1)$$

The speed of sound $c_s(t)$ in terms of the speed of light $c(t)$ in the photon-baryon fluid with baryon density $\Omega_b$ and radiation density $\Omega_r$ is given by (Durrer 2008)

$$c_s(t) \approx \frac{c(t)}{\sqrt{3}} \left(1 + \frac{3\Omega_b}{4\Omega_r}\right)^{-1/2}. \qquad (B2)$$

Substituting it in equation (B1) and making use of equation (A26), we get

$$d_{sh}(z_{ls}) = \frac{c_0}{\sqrt{3}H_0(1+z_{ls})} \int_\infty^{z_{ls}} \frac{\exp[((1+z)^{-\alpha}-1)]dz}{\left(1+\frac{3\Omega_b}{4\Omega_r}\right)^{1/2} E(z)}. \qquad (B3)$$

This distance represents the maximum distance over which the baryon oscillations imprint maximum thermal radiation fluctuations observed as anisotropies in the CMB power spectrum. Such distance characterizes an angular size $\theta_{sh}$ observed at an angular diameter distance $d_A(z_{ls})$ given simply by

$$D_A(z_{ls}) \equiv S_k(d_P)/(1+z_{ls}), \text{ and } \theta_{sh} \equiv d_{sh}(z_{ls})/D_A(z_{ls}). \qquad (B4)$$

The corresponding multipole moment is given by $l_e = \pi/\theta_{sh}$. However, we need to implement a correction to the angular diameter distance due to time dilation resulting from variation in the speed of light. It is similar to the time dilation in calculating the luminosity distance. As per the discussion following equation (14), one can see that we need to change $(1+z_{ls})$ to $(1+z_{lz})f^{1/2}$ in equation (B4) where function $f$ is evaluated at $z = z_{ls}$. Since CMB originates at the surface of the last scattering, we have $z_{ls} \equiv z_{CMB}$.

The next thing for us to see is how the redshift of the last scattering surface $z_{ls}$ is modified in the VPC model. The conservation of the number of photons leads to the evolution of the energy density of the photons. The number density of photons is

$$n = n_0 a^{-3}. \tag{B5}$$

As already discussed following equation (12), the energy of a photon:

$$E_{ph} \equiv \frac{hc}{\lambda} = \frac{h_0 f c_0 f}{\lambda_0 a} = \frac{h_0 c_0 f^2 a^{-1}}{\lambda_0} = E_{ph,0} f^2 a^{-1}. \tag{B6}$$

Therefore, we get the energy density of the photons as

$$\varepsilon_{ph} \equiv E_{ph} n = E_{ph,0} f^2 a^{-1} n_0 a^{-3} = (E_{ph,0} n_0) f^2 a^{-4} \equiv \varepsilon_{ph,0} a^{-4} f^2 \tag{B7}$$

Now the blackbody radiation energy density is given by (Ryden 2017)

$$\varepsilon_\gamma = \frac{8\pi^5 k_B^4 T^4}{15 h^3 c^3} = \frac{8\pi^5 k_{B,}^4 T^4}{15 h_0^3 c_0^3 f^6} \tag{B9}$$

For it to be consistent with equation (B7), we must have $k_B T \sim f^2 a^{-1}$. Now for $\Lambda$CDM model $f = 1$ ($\alpha = 0$), and $k_B$ does not evolve, and therefore we get $T \sim a^{-1}$, as expected. However, when $\alpha \neq 0$, we have to consider $k_B T = k_{B,0} T_0 f^2 a^{-1} \equiv k_{B,0} T_0 a'^{-1}$, where $k_{B,0}$ is the present value of the Boltzmann constant, and $T_0$ is the current CMB temperature (2.7255 K). This means the energy of the emitted CMB photons at the last scattering surface is lower by a factor of $f^2$ (for $\alpha > 0$) in the VPC model than in the $\Lambda$CDM model. Since the Rydberg unit of energy, which relates to the atomic energy levels and the ionization energies, also scales as $f^2$ [equation (11): $R_y \sim f^2$], ionization energy is also a factor $f^2$ lower in the VPC model than in the $\Lambda$CDM model at the last scattering surface. Therefore, we may write

$$a'^{-1} \equiv f^2 a^{-1} \Rightarrow 1 + z' \equiv f^2(1 + z) \equiv 1 + z_{CMB}. \tag{B10}$$

We could also see it by studying how the form of the Planck blackbody radiation spectrum evolves in the VPC model. The intensity of the blackbody radiation per wavelength interval, $B_\lambda d\lambda$, is given by (Maoz 2016)

$$B_\lambda d\lambda = \frac{2hc^2}{\lambda^5} \left(\frac{d\lambda}{\exp(hc/\lambda k_B T) - 1}\right). \tag{B11}$$

Dividing by the energy of a photon, $ch/\lambda$, we get the number density $n$,

$$n = \frac{2c}{\lambda^4} \left(\frac{d\lambda}{\exp(hc/\lambda k_B T) - 1}\right). \tag{B12}$$

Therefore, the current number density, $n_0$, maybe written as

$$n_0 = \frac{2c_0}{\lambda_0^4} \left(\frac{d\lambda_0}{\exp(h_0 c_0/\lambda_0 k_{B,0} T_0) - 1}\right). \tag{B13}$$

Since the number density $n$ scales as $a^{-3}$ and $\lambda$ scales as $a$, we may write the photon number density, $n_e$, at the time of CMB emission in terms of $n_0$ as

$$n_e = n_0 a_e^{-3} = \frac{2c_0}{\lambda_0^4 a_e^4} \left(\frac{d\lambda_0 a_e}{\exp(a_e h_0 c_0 f_e^2/\lambda_0 a_e k_{B,0} T_0 f_e^2) - 1}\right), \text{ or}$$

$$n_e = \frac{2c_0 f_e}{\lambda_e^4 f_e} \left(\frac{d\lambda_e}{\exp(h_e c_e/\lambda_e k_{B,0} T_0 a_e^{-1} f_e^2) - 1}\right) = \frac{2c_e}{\lambda_e^4 f_e} \left(\frac{d\lambda_e}{\exp[h_e c_e/\lambda_e (k_B T)_e] - 1}\right). \tag{B14}$$

Here we have defined the thermal energy parameter $(k_BT)_e \equiv k_{B,0}T_0 f_e^2 a_e^{-1} \equiv k_{B,e}T_e a_e^{-1} = k_{B,e}T_e(1+z_e)$. It shows that the form of the Planck spectrum remains unchanged under the VPC approach, and the photon energy evolves with an extra factor $f_e^2$.

Next, we must see if the above treatment is consistent with the energy continuity equation (Gupta 2020)

$$\dot{\varepsilon} + 3\frac{\dot{a}}{a}(\varepsilon + p) = 0. \tag{B15}$$

From equation (B7), we may write for photons

$$\varepsilon = \varepsilon_0 a^{-4} f^2 \Rightarrow \dot{\varepsilon} = -4\varepsilon_0 a^{-5}\dot{a}f^2 + 2\varepsilon_0 a^{-4} f\dot{f}. \tag{B16}$$

But,

$$f = \exp(a^\alpha - 1) \Rightarrow \dot{f} = \exp(a^\alpha - 1)\alpha a^{\alpha-1}\dot{a} = f\alpha a^\alpha \frac{\dot{a}}{a}. \tag{B17}$$

Therefore,

$$\dot{\varepsilon} = -4\varepsilon_0 a^{-4}\frac{\dot{a}}{a}f^2 + 2\varepsilon_0 a^{-4}f^2\alpha a^\alpha \frac{\dot{a}}{a} = -4\varepsilon_0 a^{-4}\frac{\dot{a}}{a}f^2(1 - \frac{1}{2}\alpha a^\alpha). \tag{B18}$$

Case 1: The LCDM model, i.e., $\alpha = 0$. Then, $f = 1$ and $\dot{\varepsilon} = -4\varepsilon_0 a^{-4}(\dot{a}/a)$ and equation (B15) becomes

$$-4\varepsilon_0 a^{-4}\frac{\dot{a}}{a} + 3\frac{\dot{a}}{a}(\varepsilon_0 a^{-4} + p) = 0. \tag{B19}$$

This yields the correct equation state for photons:

$$p = \frac{1}{3}\varepsilon_0 a^{-4} = \frac{1}{3}\varepsilon. \tag{B20}$$

Case 2: The VPC model, i.e., $\alpha \neq 0$. Let us write for brevity $B \equiv (1 - \frac{1}{2}\alpha a^\alpha)$. Then, equation (B18) is

$$\dot{\varepsilon} = -4\varepsilon_0 a^{-4}\frac{\dot{a}}{a}f^2 B \tag{B21}$$

Equation (B15) is now

$$-4\varepsilon_0 a^{-4}\frac{\dot{a}}{a}f^2 B + 3\frac{\dot{a}}{a}(\varepsilon_0 a^{-4}f^2 + p) = 0, \text{ or} \tag{B22}$$

$$(4B - 3)\varepsilon_0 a^{-4}f^2 = 3p \Rightarrow p = \frac{1}{3}\varepsilon_0 a^{-4}f^2(4B - 3) = \frac{1}{3}(4B - 3)\varepsilon. \tag{B23}$$

Substituting for B, the equation of state for this case is

$$p = \frac{1}{3}(4 - 2\alpha a^\alpha - 3)\varepsilon = \frac{1}{3}(1 - 2\alpha a^\alpha)\varepsilon \equiv w(a,\alpha)\varepsilon. \tag{B24}$$

This yields $w = 1/3$ for $\alpha = 0$ as expected. The same value, $w = 1/3$ is true for $a \ll 1$ with all plausible $\alpha$ values. Since the radiation dominant era ended at $a \approx 0.0003$, $w$ can be taken as 1/3 all through that era. Even in the matter dominated era, $w$ was close to 1/3 until $\Omega_r \ll \Omega_m$. However, $w$ becomes negative when $2\alpha a^\alpha$ becomes greater than 1, e.g., at $a \approx 0.5$ for $\alpha = 1.8$ (our preferred value). But at such high values, $\Omega_r \ll \Omega_m$ and, therefore, the negative value region of $w$ for radiation is meaningless. Thus, for all practical purposes, $w = 1/3$ for radiation in our model.

*BAO acoustic scale*: Baryonic acoustic oscillations are associated with the CMB temperature anisotropy. Since the galaxies are believed to have their origin imprinted on the CMB perturbations, their scale is considered measurable today through the correlation function representing the distribution of galaxies in space (Anderson et al. 2014). In the ΛCDM model, the acoustic scale is expressed as

$$r_{as} = d_{sh}(z_{ls})(1 + z_{ls}). \tag{B25}$$

This expression needs to be corrected for the VPC model: (i) Time dilation due to varying speed of light adds a factor $f^{1/2}$ to the right-hand side of equation (B11) as we did for the calculation of $l$ above. (ii) We have to correct it for the curvature of the Universe from the observer ($z = 0$) to the galaxies used in determining the correlation function ($z = 0.2$ to $0.7$). Because most galaxy counts would be at higher redshift and distances are larger in the VPC model (equation 19), we have taken $z \approx 0.7$ for making this correction. Correction is then obtained by evaluating the ratio of the proper distance, $d_{p,c}$, of the surface of the last scattering from $z = 0.7$ to $z = z_{CMB}$ with corresponding $S_k(d_{p,c})$: Thus, we define the observed acoustic scale as $A_s \equiv r_{as}(d_{p,c}/S_k(d_{p,c}))$. In a positively curved universe, we observe $r_{as}$ larger than it actually is and smaller in the negatively curved Universe.

*Deceleration parameter:* For the VPC model, it may be written as (Gupta 2020),

$$q_0 = \tfrac{1}{2}\Omega_{m,0} + \Omega_{r,0} - \Omega_{\Lambda,0} - \alpha. \tag{B26}$$

It differs from the ΛCDM expression for $q_0$ by the last term.

*Age of the Universe:* This is the current cosmic time, $t_0$, evaluated relative to the big-bang time. From equation (A18), we have

$$\frac{H}{H_0} = E(a) \Rightarrow \frac{\dot{a}}{a} = H_0 E(a) \Rightarrow \frac{da}{dt} = H_0 a E(a) \Rightarrow H_0 dt = \frac{da}{aE(a)}, \text{ and} \tag{B27}$$

$$t_0 = \int_0^{t_0} dt = \frac{1}{H_0}\int_0^1 da/aE(a). \tag{B28}$$